\documentclass[letterpaper,twocolumn]{jpsj3}
\usepackage{txfonts}
\usepackage{graphicx}
\usepackage{amsmath}
\usepackage{amsmath,amssymb}
\usepackage{bm}
\usepackage[usenames]{color}
\title{
Photoinduced Bidirectional Magnetism against Monodirectional Electronics\\
in Square-Antiprismatic Octacyanometalates}
\author{Jun Ohara and Shoji Yamamoto\thanks{yamamoto@phys.sci.hokudai.ac.jp}}
\inst{Department of Physics, Hokkaido University, Sapporo 060-0810, Japan}

\abst{
Irradiating ${\rm Cu}_2{\rm Mo}({\rm CN})_8\cdot 8{\rm H}_2{\rm O}$
with blue light induces a global magnetization,
whereas succeeding irradiations with red or longer-wavelength light
demagnetize this material.
We solve the time-dependent Schr\"odinger equation
for an extended Hubbard model
to reproduce the photoreversible magnetism.
Monitoring the photoinduced optical-conductivity
and angle-resolved-photoemission spectra,
we reveal that the magnetic round trip by way of
ferromagnetism is far from a return in terms of electronics.
While visible-light-induced magnetization has never been observed
in the tungsten analog ${\rm Cu}_2{\rm W}({\rm CN})_8\cdot 5{\rm H}_2{\rm O}$,
infrared-light irradiation may magnetize this material as well.}

\begin{document}
\maketitle
\begin{figure}[ht]
\centering
\includegraphics[width=85mm]{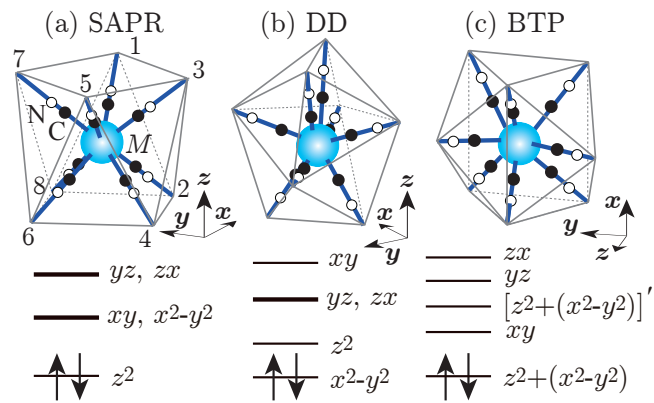}
\vspace*{-3mm}
\caption{(Color online)
Spatial configurations of [$M^{\rm IV}$(CN)$_{8}$]$^{4-}$ ($M$=Mo, W):
(a) Square antiprism (SAPR), (b) dodecahedron (DD),
and (c) bicapped trigonal prism (BTP).
The crystal-field splittings of the $4d$ ($5d$) orbitals on
the Mo (W) site and the electron configurations are also shown,
where thick lines denote doubly degenerate levels.}
\label{F:UNIT}
\end{figure}

    Multifunctional magnetic materials
based on octacyanometalates \cite{K1087,L215}
have been attracting considerable attention.
In comparison with popular hexacyanometalates
including Prussian blue analogs \cite{S704,M1023}
well known to be spin-crossover complexes,
higher-than-six-coordinate metal complexes
can have various magnetic exchange pathways \cite{S2203,P2234}.
The octacyano transition-metal ions
[$M$(CN)$_{8}$]$^{3-/4-}$ ($M$=Mo, W)
can take spatial configurations such as
square antiprism (SAPR) of ${\bf D}_{4d}$ point symmetry [Fig. \ref{F:UNIT}(a)],
dodecahedron (DD) of ${\bf D}_{2d}$ point symmetry [Fig. \ref{F:UNIT}(b)],
and bicapped trigonal prism (BTP) of ${\bf C}_{2v}$ point symmetry
[Fig. \ref{F:UNIT}(c)] \cite{P2556,B2553,M109}.
Among others, the coordination networks comprising
the square-antiprismatic anions [Mo$^{\rm IV}$(CN)$_{8}$]$^{4-}$ 
and cations Cu$^{2+}$ are so interesting
as to exhibit photoswitchable magnetism \cite{R27,O523,M094107}.
Blue-light irradiation of the dicopper octacyanomolybdate
${\rm Cu}_2{\rm Mo}({\rm CN})_8\cdot8{\rm H}_{2}{\rm O}$,
abbreviated as Cu-Mo, at sufficiently low temperatures
induces ferromagnetism,
whereas subsequent longer-wavelength-light irradiation
demagnetizes this material \cite{O270}.
However, little is known about the photoexcited electronic state
from the experimental viewpoint and we are almost ignorant
of the photoinduced demagnetization as well as
magnetization mechanisms.

\begin{figure}[h]
\centering
\includegraphics[width=85mm]{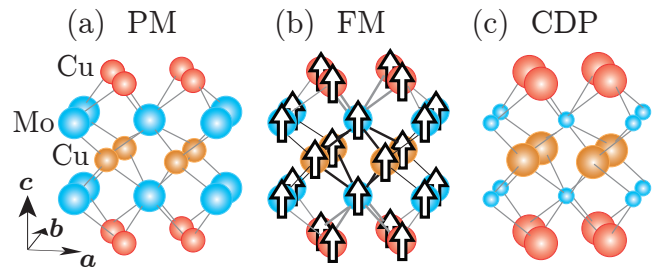}
\vspace*{-3mm}
\caption{(Color online)
(a) Paramagnetic (PM), (b) ferromagnetic (FM),
and (c) charge-disproportionated (CDP) states
in the dicopper octacyanomolybdate (Cu-Mo),
where varying circles and arrows describe
varying electron and spin densities, respectively.
The crystal axes are denoted by
${\bm a}$, ${\bm b}$, and ${\bm c}$.}
\label{F:PHASE}
\end{figure}
    In the $I4/m$ crystal structure
of ${\bf C}_{4h}$ point symmetry \cite{SC},
the constituent Mo$^{\rm IV}$ and Cu$^{\rm II}$ ions
contribute their doubly occupied 4$d_{z^{2}}$
and singly occupied 3$d_{zx/yz}$ orbitals
to the itinerant magnetism, respectively.
The 3$d_{zx}$ (3$d_{yz}$) orbitals are sitting
on the Cu ions located at the vertices 1, 2, 5, and 6
(3, 4, 7, and 8) of SAPR [Fig. \ref{F:UNIT}(a)]  \cite{SOB}.
We employ a multiband extended Hubbard
model \cite{O17006,Y044709,SH}
to visualize the photoexcitation mechanism theoretically
and encourage further explorations of related materials.
The Hubbard Hamiltonian is treated
within a Hartree-Fock (HF) scheme,
and it is revealed that the mixed-valent states
shown in Fig. \ref{F:PHASE}
compete with each other at
sufficiently low temperatures. \cite{SP}
The paramagnetic (PM) state
can take the valence arrangement
Cu$^{\rm II}$-Mo$^{\rm IV}$-Cu$^{\rm II}$
and appears as the ground state
unless the electronic correlation is not
sufficiently strong.
In a strong correlation regime,
the ferromagnetic (FM) state having a global magnetization
is stable.
This state shows a spatially uniform electron distribution
and exhibits metallic behavior.
A non-magnetized solution different from the PM state,
the charge-disproportionated (CDP) state,
appears in Cu-Mo.
The CDP state takes the valence arrangement
Cu$^{\rm I}$-Mo$^{\rm VI}$-Cu$^{\rm I}$
(fully-filled Cu and vacant Mo sites)
and plays an important role in the photoexcited Cu-Mo.
The model parameters are tuned so as to
reproduce the valence arrangement Cu$^{\rm II}$-Mo$^{\rm IV}$-Cu$^{\rm II}$
in the PM ground state. \cite{SH}

    We consider Dzyaloshinskii-Moriya (DM) and Zeeman interactions
in a time-evolutional calculation.
The DM interaction arises from the crystalline structure \cite{S775,K3946,SD},
which allows the total magnetization to fluctuate.
We apply a weak external magnetic field
along the ${\bm c}$ (${\bm z}$) axis
in line with the experimental setup \cite{O270}.
Though the Zeeman interaction \cite{SZ}
serves to lift the spin up-down degeneracy,
the applied field is sufficiently weak and
can never justify in itself the observed giant
photomagnetization.
We introduce the vector potential ${\bm A}(t)$
to photoirradiate the system, i.e.,
multiply the electron hopping term
by the Peierls phase factor
$\displaystyle \exp[-\frac{{\rm i}e}{\hbar v}{\bm A}(t)\cdot{\bm \delta}]$, \cite{T177}
where $e$, $v$, and ${\bm \delta}$
are the elementary charge, speed of light,
and relative vector between
cyano-bridged metals, respectively.
Time evolution of an electronic wavefunction is obtained
by integrating the time-dependent Schr\"{o}dinger equation
$i\hbar|\dot{\Psi}(t)\rangle={\cal H}_{\rm HF}(t)|\Psi(t)\rangle$,
where ${\cal H}_{\rm HF}(t)$ is
the total Hamiltonian treated within the HF scheme \cite{T024712,Y084713}.
At $t\to-\infty$,
the wavefunction $|\Psi(t)\rangle$ corresponds to the initial PM state.
Hereafter, 
the transfer integral between the nearest neighboring
cyano-bridged metals $t_{\rm MoCu}$ is set to $0.65$ eV,
which is comparable to the transfer integral in
typical Prussian blue analogs \cite{K12990,Y5432}.

\begin{figure}[t]
\centering
\includegraphics[width=74mm]{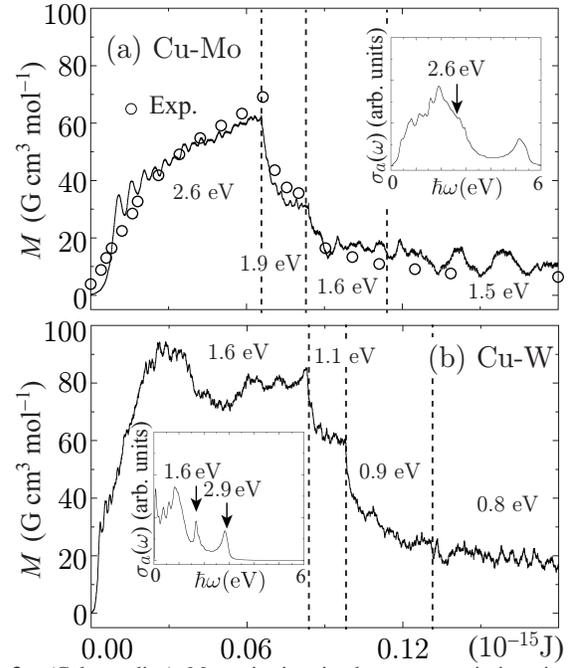}
\vspace*{-5mm}
\caption{(Color online)
Magnetization in the square-antiprismatic octacyanometalates
due to successive photoirradiation plotted as
a function of the average energy absorbed by each unit cell,
calculations (solid lines) and experiments (open circles).\cite{O270}
(a) Cu-Mo: the photon energy, duration, and amplitude
are taken as $\hbar\omega_{i}=2.6, 1.9, 1.6, 1.5$ eV,
$\tau_{i}=0.9, 0.3, 0.3, 0.5$ ps,
and $|\bm{A}_{i}|ea/\hbar v=0.5, 0.6, 1.0, 1.1$,
respectively.
The switching time $2\gamma^{-1}$ is set to 0.03 ps.
(b) Cu-W: the same as (a) but $\hbar\omega_{i}=1.6, 1.1, 0.9, 0.8$ eV.
Insets in (a) and (b) show the polarized ($\parallel{\bm a}$)
optical conductivities in the ground state.}
\label{F:FIT}
\end{figure}

    Figure \ref{F:FIT}(a) shows the magnetization dynamics in Cu-Mo
induced by continuous laser lights polarized parallel to ${\bm a}$ axis,
\begin{align}
{\bm A}(t)=\sum_{i=1}^{4}{\bm A}_{i}&
\frac{[1+\tanh\{ \gamma (t-t_{i}^{\rm on})\}]}{2}
\frac{[1+\tanh\{-\gamma (t-t_{i}^{\rm off})\}]}{2}
\nonumber \\ \times &
\cos{\omega_{i}t}
\label{E:CW}
\end{align}
with various frequencies $\omega_{i}$ of duration
$\tau_{i} \equiv t_{i}^{\rm off}-t_{i}^{\rm on}$.
The power and duration of the laser lights are on the order of
$10^{12}$ J s$^{-1}$cm$^{-2}$ and $10^{-12}$ s,
respectively.
They are standard in photoinduced-dynamics calculations \cite{Y045134}
but not necessarily comparable with experimental observations. \cite{O270}
So, we simulate and reproduce any experiment
based on the average energy absorbed by each unit cell
during the light irradiation.
We assume that effects of energy dissipation
are negligible in the magnetization dynamics
\cite{O523,SRL}.
The irradiation with $\hbar\omega=2.6$ ${\rm eV}$ light
induces a global magnetization,
whereas the subsequent irradiation with $\hbar\omega=1.9$ ${\rm eV}$ light
conversely demagnetizes the system.
The magnetization further decreases gradually
upon the third and fourth irradiations with longer-wavelength light.
The simulations well reproduce the visible-light-induced
magnetization and demagnetization dynamics
in the experimental observation \cite{O270}.
Details of the electronic structures
are discussed later.
We further investigate photoinduced dynamics
in the tungsten analog complex
${\rm Cu}_2{\rm W}({\rm CN})_8\cdot5{\rm H}_{2}{\rm O}$ \cite{R27},
abbreviated as Cu-W.
Its ground state shows the valence arrangement
Cu$^{\rm II}$-W$^{\rm IV}$-Cu$^{\rm II}$,
where the W$^{\rm IV}$ ions contribute 
their doubly occupied $5d_{z^{2}}$ orbitals [Fig. \ref{F:UNIT}(a)]
to the physical properties.
This situation is qualitatively the same as the PM state in Cu-Mo.
However, irradiations of the ground state
with $\hbar\omega=$2.5 and 3.0 eV lights
induce no significant magnetization \cite{R27}.
According to these observations,
Cu-W seems not to be a photomagnet.
We set the transfer integral $t_{\rm WCu}$ to 0.42 eV
($< t_{\rm MoCu}$) by considering metal-substitution effects \cite{ST}
and irradiate the ground state with the continuous laser light.
The violet-light ($\hbar\omega=$2.9 eV) irradiation induces no global magnetization,
but the sufficiently low-energy-light
($\hbar\omega=$1.6 eV) irradiation causes a PM-to-FM transition [Fig. \ref{F:FIT}(b)].
The magnetization induced by this infrared light disappears
with succeeding irradiations.
Thus, Cu-W, as well as Cu-Mo, shows photoswitchable magnetism.
The insets in Fig. \ref{F:FIT} show
the ground-state optical conductivities
along ${\bm a}$ axis \cite{W10517,S2830,SO}.
A major absorption band of Cu-W is observed
in an infrared region,
while that of Cu-Mo is seen in a visible region.
The electron-hole pair generation by
these optical absorptions leads to the magnetic transitions.
Note that these ground-state absorption spectra are also
in good agreement with the observations \cite{O270,R27}.

\begin{figure}[ht]
\centering
\includegraphics[width=80mm]{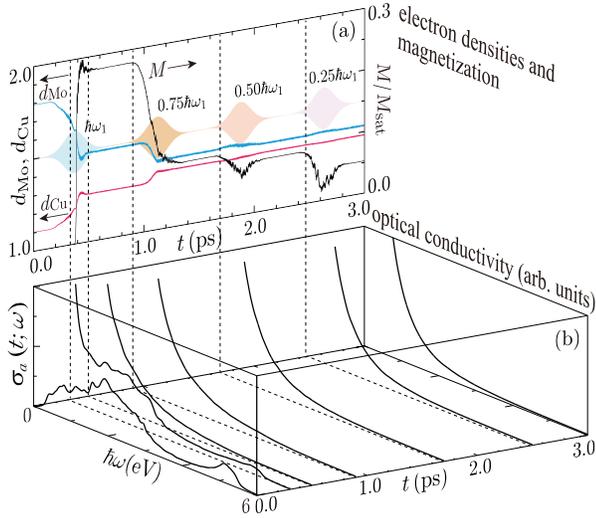}
\vspace*{-4mm}
\caption{(Color online)
Photoinduced magnetization and demagnetization dynamics:
(a) The averaged electron density on Mo (Cu) sites $d_{\rm Mo\,(Cu)}$
and the total magnetization divided by the saturation magnetization $M/M_{\rm sat}$.
The pumping laser pulses are such that
$\hbar\omega_{i}=2.6[1-0.25(i-1)]$ eV
and $t_{i}=0.4+0.75(i-1)$ ps,
illustrated schematically in (a).
The amplitude $|\bm{A}|ea/\hbar v$ and width $2\gamma^{-1}$ of each pulse
are set to 2.0 and 0.25 ps, respectively.
(b) Photoinduced optical conductivities
along ${\bm a}$ axis at several stages in
the magnetization and demagnetization processes.}
\label{F:MD}
\end{figure}

    Figure \ref{F:MD}(a) shows the magnetization dynamics in Cu-Mo
induced by pulsed laser lights polarized parallel to ${\bm a}$ axis,
\begin{align}
{\bm A}(t)={\bm A}\sum_{i=1}^{4}
e^{-\gamma^{2}(t-t_{i})^{2}}\cos{\omega_{i}t}
\end{align}
with various frequencies $\omega_{i}$ of width $2\gamma^{-1}$.
In order to monitor the magnetic dynamics in more detail,
we calculate the electron densities on transition-metal sites
and the total magnetization as functions of time.
The irradiation with $\hbar\omega_{1}$ light
firstly causes the electron transfer from Mo to Cu sites
(0.2 $\sim$ 0.3 ps) and secondary induces
the substantial magnetization (0.4 ps).
Since the spin degeneracy is
lifted due to the magnetic field,
the higher-lying down-spin electrons are
selectively excited, \cite{SE}
which leads to a net spin density on each site.
After that, the DM interaction begins to work.
The magnetization switching speed increases in proportion to
the increase in the laser amplitude $|{\bm A}|$. \cite{SAH}
In this FM transition,
metallization proceeds the magnetization
as is demonstrated by the optical conductivities
in Fig. \ref{F:MD}(b),
where the Drude response is observed.
The magnetization process reads as
an insulator-to-metal transition.
The subsequent irradiation with $0.75\hbar\omega_{1}$ light
demagnetizes the system (0.9 $\sim$ 1.3 ps),
while electrons are further transferred from Mo to Cu sites
and the conductivity does not change.
This is not just a reverse process of the photomagnetization.
When the system is further irradiated with
$0.50\hbar\omega_{1}$ and $0.25\hbar\omega_{1}$ lights successively,
the magnetization gets small step by step. \cite{SClm}
The transferred electrons never turn back to Mo sites
and the system remains metallic.
The final demagnetized state is totally different
from the initial insulating PM state.
Qualitatively the same phase transitions occur
when the system is irradiated with
the continuous laser lights (Fig. \ref{F:FIT}).
Thus, the magnetic round trip is far from a return
in terms of the electric conductivity.

\begin{figure}[t]
\centering
\includegraphics[width=86mm]{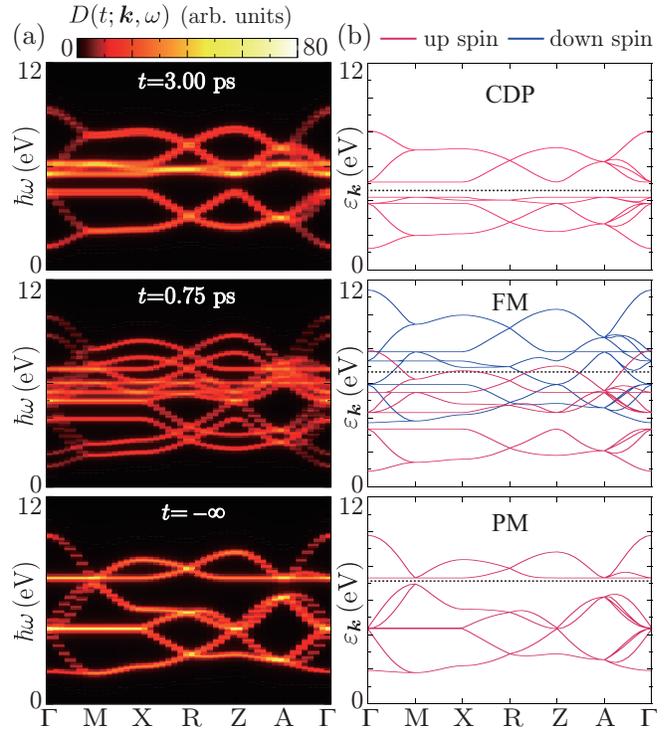}
\vspace*{-7mm}
\caption{(Color online)
(a) ARPES spectra of
the initial PM ($t=-\infty$), 
photomagnetized ($t=0.75$ ps),
and final demagnetized ($t=3.0$ ps) states.
(b) Dispersion relations of up- and down-spin electrons
in the PM, FM, and CDP ground states,
where a dotted line in each panel denotes the Fermi level.}
\label{F:ARPS}
\end{figure}

    Figure \ref{F:ARPS}(a) shows
the one-particle excitation spectrum $D(t;{\bm k},\omega)$
corresponding to an angle-resolved
photoemission spectroscopy (ARPES) spectrum \cite{K045137}:
$D(t;{\bm k},\omega)=H(t;{\bm k},\omega)+E(t;{\bm k},\omega)$
with
\begin{align}
E(t;{\bm k},\omega)
&=\sum_{{\bm k},\chi}
\delta(\hbar\omega-\varepsilon_{{\bm k},\chi})
\langle{\Psi(t)}|
\tilde{c}_{{\bm k},\chi}^{\dagger}\tilde{c}_{{\bm k},\chi}
|{\Psi(t)}\rangle,\\
H(t;{\bm k},\omega)
&=\sum_{{\bm k},\chi}
\delta(\hbar\omega-\varepsilon_{{\bm k},\chi})
\langle{\Psi(t)}|
\tilde{c}_{{\bm k},\chi}\tilde{c}_{{\bm k},\chi}^{\dagger}
|{\Psi(t)}\rangle,
\end{align}
where $\varepsilon_{{\bm k},\chi}$ is
a HF-level energy of ${\cal H}_{\rm HF}(t)$
and $\tilde{c}_{{\bm k},\chi}^{\dagger}$ creates
the electron labeled by the momentum ${\bm k}$
and the band index $\chi$ $(=1\sim24)$.
We compare band structures of the photoexcited system
with those of the PM, FM, and CDP ground states
[Fig. \ref{F:ARPS}(b)].
We can obtain each ground state by
tuning the on-site Coulomb repulsion $U_{\rm Cu/Mo}$. \cite{SP}
The first irradiation with $\hbar\omega=2.6$ ${\rm eV}$ light excites
electrons around $\Gamma$ point.
The spatially uniform oscillation
leads to the appearance of the macroscopic magnetization.
In the photomagnetized metallic state ($t=0.75$ ps),
the spin degeneracy is totally lifted just like
the itinerant FM ground state.
In the photoexcited state,
the spin index up or down is
no longer a good quantum index due to the DM interaction.
Further photoirradiation gives rise to
the charge-transfer excitations accompanied by spin flips,
which leads to the reduction in
the net spin density on each site.
It is observed as convergence of the split bands.
The band structure of the thus demagnetized state ($t=3.00$ ps)
is very close to that of the CDP ground state.
The static CDP structure is stable
in the condition $U_{\rm Mo} \gg U_{\rm Cu}$ \cite{SP},
but this situation is hardly realized
because Mo 4$d$ orbitals are more diffuse than Cu 3$d$ ones
in general.
Such a hidden structure appears in the photoexcited Cu-Mo
and the final steady state should be called
the {\it nonequilibrium} CDP state.
Note that further light irradiation
of excited states leads to emission
as well as absorption of photons.
Photoexcited electrons partially
turn back to the valence bands.
Since a resonant energy varies with changes
in the band structure [Fig. \ref{F:ARPS}(a)],
emergence of the bidirectional magnetism
requires the successive irradiation
with varying wavelength light.

\begin{figure}[ht]
\centering
\includegraphics[width=86mm]{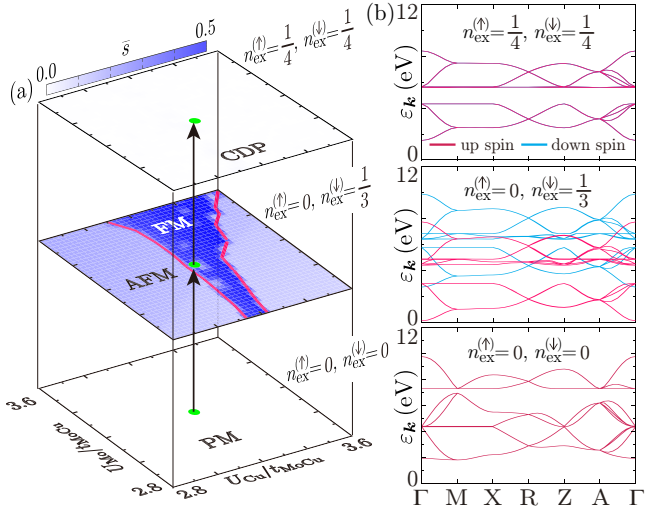}
\vspace*{-7mm}
\caption{(Color online)
(a) Phase competition in the ground and photoexcited states
on the $U_{\rm Cu}$-$U_{\rm Mo}$ plane,
where the averaged spin density $\bar{s}$
$(=|{\bm s}_{\rm Mo}|/3+2|{\bm s}_{\rm Cu}|/3)$ is also shown.
(b) Band structures of the ground and photoexcited states at
$U_{\rm Mo}=U_{\rm Cu}=3.2t_{\rm MoCu}$.}
\label{F:PDEX}
\end{figure}

    Figure \ref{F:PDEX}(a) shows
the phase competition in the ground and excited states.
We tune the electron population of the bands
to simulate the photoexcitation,
where the population is measured by
numbers of excited up-spin ($\uparrow$)
and down-spin ($\downarrow$) electrons per site,
$n^{(\uparrow)}_{\rm ex}$ and $n^{(\downarrow)}_{\rm ex}$.
As is shown in Fig. \ref{F:PDEX}(a),
we make phase diagrams of the static
Hubbard Hamiltonian under several conditions.
We focus on the region around
the present Coulomb parameters
$U_{\rm Mo}=U_{\rm Cu}=3.2t_{\rm MoCu}$, 
in which the PM ground state predominates.
For example, the condition
$n^{(\uparrow)}_{\rm ex}=0$ and $n^{(\downarrow)}_{\rm ex}=1/3$
means that all of the up-spin electrons are in valence bands
but half of the down-spin ones in valence bands
are excited to conduction bands.
On the basis of these electron distributions,
we calculate order parameters,
such as electron and spin densities,
and then obtain the mean-field Hamiltonian
describing photoexcited states.
Diagonalizing this Hamiltonian,
we find an energy and band structure of the excited state.
We simulate the first photoexcitation
by considering the condition
$n^{(\uparrow)}_{\rm ex}=0$ and
$n^{(\downarrow)}_{\rm ex}=1/3$,
which describes the spin-selective electron excitation.
The FM and antiferromagnetic (AFM) ordering
between the spins on Mo and Cu sites
compete with each other.
At the present parameter point,
the FM structure appears with
the electron-hole pair excitations.
Considering the spin mixing and photon emission
upon further light irradiations,
we take the condition $n^{(\uparrow)}_{\rm ex}=
n^{(\downarrow)}_{\rm ex}=1/4$
as the final steady state.
The CDP structure becomes more stable
than the PM one under this nonequilibrium situation.
Figure \ref{F:PDEX}(b) shows
the band structure under each condition.
Its $n^{(\sigma)}_{\rm ex}\,(\sigma=\uparrow,\downarrow)$
dependence is reminiscent of
the time evolution of the photoexcited state
shown in Fig. \ref{F:ARPS}(a).
Thus, the bidirectional photoswitching of magnetization
reads as the multistep photoinduced phase transition.

    Cobalt octacyanotungstates Co$^{\rm III}$-W$^{\rm IV}$ in
the DD [Fig. \ref{F:UNIT}(b)] and BTP [Fig. \ref{F:UNIT}(c)] geometries
also exhibit photoinduced ferromagnetism \cite{O2089,A9240}.
The Co$^{\rm III}$ and W$^{\rm IV}$ ions
in the ground state contribute their vacant $3d_{z^{2}}$ and
fully-filled $5d_{(x^{2}-y^{2})/[z^{2}+(x^{2}-y^{2})]}$
orbitals to the photomagnetism \cite{O2089,A9240,S37}.
Unlike the PM ground states in
Cu$^{\rm II}$-Mo$^{\rm IV}$ and Cu$^{\rm II}$-W$^{\rm IV}$,
the Coulomb repulsion between Co and W sites
stabilizes the valence arrangement Co$^{\rm III}$-W$^{\rm IV}$.
The observed magnetization is about an order of magnitude larger
than that in the SAPR systems.
Four Co ions surround each W ion, i.e.,
half the CN groups of W(CN)$_{8}$
are not linked to Co.
Therefore, the Co-W intermetallic Coulomb repulsion is thus weaker
than the Cu-Mo one in the SAPR geometry.
The photodoped carriers are more itinerant
in such a situation,
leading to larger magnetization.
The photomagnetism owes much to
the coordination geometry.

\begin{table}[]
\caption{Theoretical evaluation of whether or how
photoswitchable magnetism occurs in
various square-antiprismatic octacyanometalates.}
\begin{tabular}{cccc}
\hline \hline
$\begin{matrix}
\text{Constituent} \\[-1mm]
\text{metals}
\end{matrix}$
\hspace{2mm}
&
$\begin{matrix}
\text{Electron} \\[-1mm]
\text{occupancy}
\end{matrix}$
\hspace{2mm}&
$\begin{matrix}
\text{Pump light} \\[-1mm]
\text{energy}
\end{matrix}$
\hspace{2mm}&
$\begin{matrix}
\text{Photoswitchable} \\[-1mm]
\text{magnetism}
\end{matrix}$
\\[1mm]
\hline
\\[-2.5mm]
  Cu$_{2}^{\rm II}$Mo$^{\rm IV}$
& 3$d^{\,9}$4$d^{\,2}$
& 1.5 $\sim$ 2.6 eV
& Yes\\[1mm]
  Cu$_{2}^{\rm II}$W$^{\rm IV}$
& 3$d^{\,9}$5$d^{\,2}$
& 0.8 $\sim$ 1.6 eV
& Yes\\[1mm]
  Ni$_{2}^{\rm II}$Mo$^{\rm IV}$
& 3$d^{\,8}$4$d^{\,2}$
& ---
& No\\[1mm]
\hline \hline
\end{tabular}
\label{T:Summary}
\end{table}

    The multistep photoinduced phase transitions
in the dicopper octacyanometalates
are not a turn-around process such as
a bidirectional phase transition \cite{K6265,Z064708}.
The visible- or infrared-light irradiation
indeed induces a turn-around magnetic change,
but it arises from one-way electronic transfer.
The electron-hole pair excitation around $\Gamma$ point,
which brings about a spatially uniform oscillation,
causes the substantial increase of the global magnetization.
Therefore, adjustment of the pump light energy is considerably important.
The optical conductivity measurement in the low energy region below 1 eV
is needed to detect the one-way electronic transfer.
We consider a band-filling dependence
of the photoswitchable magnetism.
Replacement of $3d_{yz/zx}^{\,1}$ by $3d_{yz/zx}^{\,0}$
changes the total band filling 
$\nu$ from 2/3 to 1/3.
It corresponds to substitution of Ni for Cu \cite{Z363}.
At $\nu=1/3$,
electrons on anion sites feel no
intermetallic Coulomb repulsion.
Such a situation is much stabler than
the PM state in Cu-Mo or Cu-W.
We confirmed that any photoirradiation
induces neither global nor stable magnetization in the Ni-Mo system.
As is shown in Table \ref{T:Summary},
the choice of cations, as well as pumping energy,
is decisive of photosensitivity of the system.
Photomagnetism in DD and BTP complexes,
their demagnetization mechanisms in particular,
are also an intriguing topic in the future.

The authors are grateful to S. Ohkoshi and H. Tokoro
for valuable comments and discussion.
This work was supported by the Ministry of Education,
Culture, Sports, Science and Technology of Japan.


\begin{thebibliography}{99}
\bibitem{SRL}%
We investigated interplay of electronic
and lattice degrees of freedom in this system
by employing a Peierls-Hubbard Hamiltonian.
No lattice deformation occurs in the PM, FM, CDP states
and no energy dissipation
from the electron to lattice systems
is observed in any time-evolutional calculation.
\end{thebibliography}
\end{document}


\maketitle
\section*{S1. Modeling}
\label{S:S1}
\begin{figure}[ht]
\centering
\includegraphics[width=86mm]{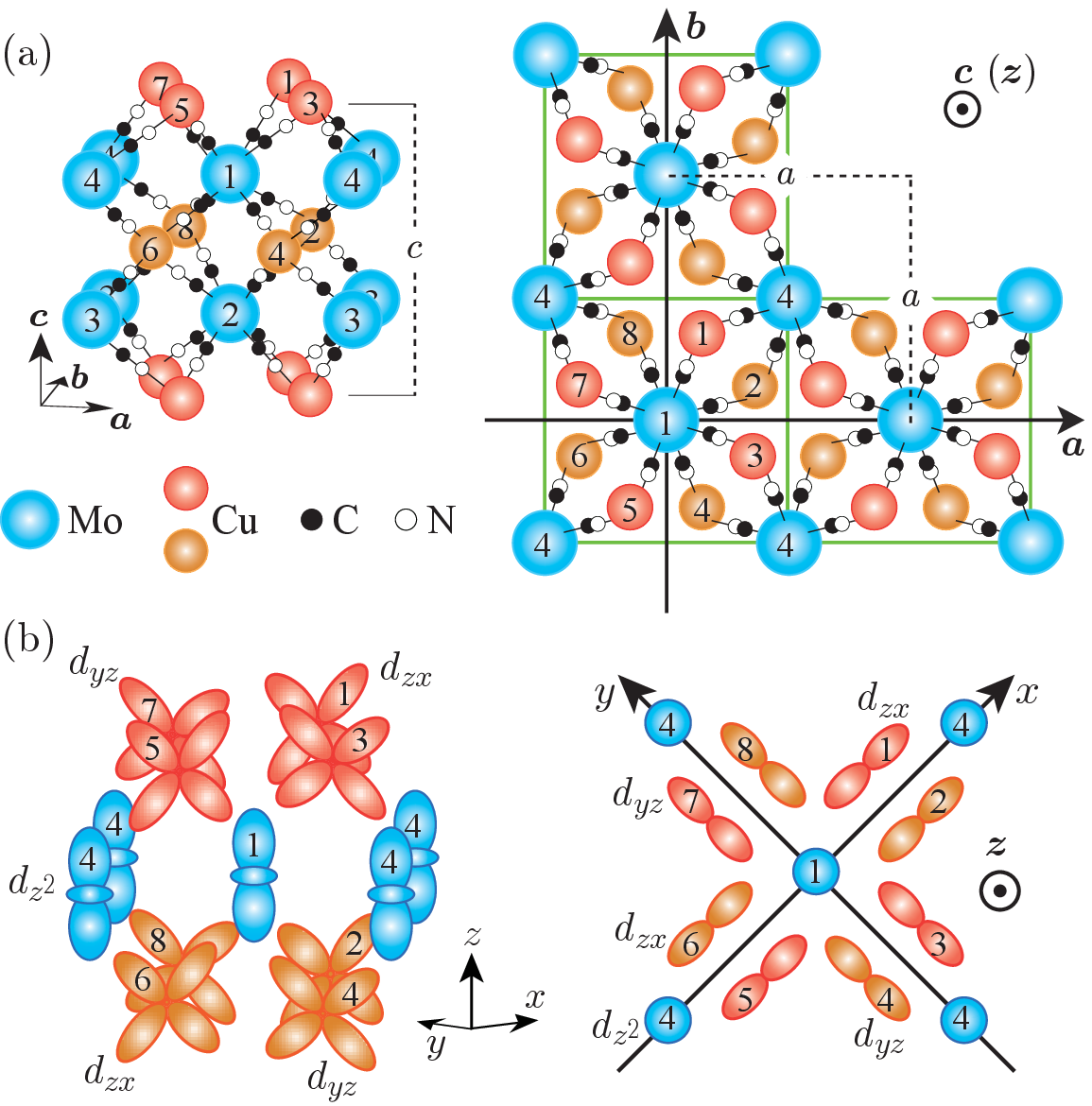}
\vspace*{-5mm}
\caption{
(a) The $I4/m$ structure of the dicopper octacyanomolybdate,
where the lattice constants $a$ and $c$ are set to 12 $\AA$. \cite{R27}.
(b) The arrangement of the Mo $d_{z^{2}}$, Cu 3$d_{zx}$, and Cu 3$d_{yz}$ orbitals.}
\label{F:S1}
\end{figure}

    To describe the supramolecular coordination network
comprising Mo$^{\rm IV}$ and Cu$^{\rm II}$ ions
[Fig. \ref{F:S1}(a)],
we employ a three-orbital extended Hubbard
Hamiltonian \cite{H705,S385},
\begin{align}
   &\!\!\!\!\!\!\!\!
   {\cal H}_{\rm H}=\sum_{{\bm n},\sigma}
   \Biggl[
    \sum_{i=1}^{8}
   \bigl( \varepsilon_{\rm Cu}n_{{\bm n}:{\rm Cu}(i)\sigma}
         +\frac{U_{\rm Cu}}{2}n_{{\bm n}:{\rm Cu}(i)\sigma}
                              n_{{\bm n}:{\rm Cu}(i)-\sigma} \bigr)
   \nonumber \\
   &
   +\sum_{j=1}^{4}
   \bigl( \varepsilon_{\rm Mo}n_{{\bm n}:{\rm Mo}(j)\sigma}
         +\frac{U_{\rm Mo}}{2}n_{{\bm n}:{\rm Mo}(j)\sigma}
                              n_{{\bm n}:{\rm Mo}(j)-\sigma} \bigr)
   \Biggr]
   \nonumber \\
   &
   +\sum_{<{\bm n},{\bm m},i,j>}\sum_{\sigma,\tau}
   \Biggl\{
   V_{\rm MoCu}
          n_{{\bm n}:{\rm Cu}(i)\sigma}n_{{\bm m}:{\rm Mo}(j)\tau}
   \nonumber \\
   &
   +J_{\rm MoCu}
         c_{{\bm n}:{\rm Cu}(i)\sigma}^{\dag}
         c_{{\bm m}:{\rm Mo}(j)\tau}^{\dag}
         c_{{\bm n}:{\rm Cu}(i)\tau}
         c_{{\bm m}:{\rm Mo}(j)\sigma}
   \nonumber \\
   &
   +\frac{J_{\rm MoCu}'}{2}
   \bigl[ c_{{\bm n}:{\rm Cu}(i)\sigma}^{\dag}
          c_{{\bm n}:{\rm Cu}(i)\tau}^{\dag}
          c_{{\bm m}:{\rm Mo}(j)\tau}
          c_{{\bm m}:{\rm Mo}(j)\sigma}
   +{\rm H.c.} \bigr]
   \nonumber \\
   &
   +\frac{t_{\rm MoCu}}{2}
   \Bigl[(-1)^{i+j+1} c_{{\bm n}:{\rm Cu}(i)\sigma}^{\dag}
                      c_{{\bm m}:{\rm Mo}(j)\sigma}
   +{\rm H.c.} \Bigr]
   \Biggl\}
   \nonumber \\
   &
   +\sum_{<{\bm n}\ne{\bm m}>}\sum_{i=1}^{4}\sum_{\sigma,\tau}
   \Biggl\{
   V_{\rm CuCu}n_{{\bm n}:{\rm Cu}(i)\sigma}n_{{\bm m}:{\rm Cu}(\bar{i})\tau}
   \nonumber \\
   &
   +J_{\rm CuCu}
         c_{{\bm n}:{\rm Cu}(i  )\sigma}^{\dag}
         c_{{\bm m}:{\rm Cu}(\bar{i})\tau}^{\dag}
         c_{{\bm n}:{\rm Cu}(i  )\tau}
         c_{{\bm m}:{\rm Cu}(\bar{i})\sigma}
   \nonumber \\
   &
   +\frac{J_{\rm CuCu}'}{2}
   \bigl[ c_{{\bm n}:{\rm Cu}(i  )\sigma}^{\dag}
          c_{{\bm n}:{\rm Cu}(i  )\tau}^{\dag}
          c_{{\bm m}:{\rm Cu}(\bar{i})\tau}
          c_{{\bm m}:{\rm Cu}(\bar{i})\sigma}
   \nonumber \\
   &
   +{\rm H.c.} \bigr]
   +\frac{t_{\rm CuCu}}{2}
   \bigl[c_{{\bm n}:{\rm Cu}(i)\sigma}^{\dag}c_{{\bm m}:{\rm Cu}(\bar{i})\sigma}
   +{\rm H.c.} \bigr]
   \Biggr\},
   \label{E:H}
\end{align}
where $\bar{i}=i+4$.
An electron number operator is defined as
$n_{{\bm n}:A(i)}=\sum_{\sigma}n_{{\bm n}:A(i)\sigma}
=\sum_{\sigma}c_{{\bm n}:A(i)\sigma}^{\dagger}c_{{\bm n}:A(i)\sigma}$,
where $c_{{\bm n}:A(i)\sigma}^{\dagger}$ creates the
electron of spin $\sigma=\uparrow,\downarrow\equiv\pm$ on
the Cu $3d_{zx}$ ($A={\rm Cu};\,i=1,2,5,6$),
Cu $3d_{yz}$ ($A={\rm Cu};\,i=3,4,7,8$), or
Mo $4d_{z^2}$ ($A={\rm Mo};\,i=1,2,3,4$) orbital
at unit cell ${\bm n}$
[Fig. \ref{F:S1}(b)].
Intermetallic transfer integral is denoted by $t_{\rm MoCu/CuCu}$.
The on-site Coulomb repulsion $U_{\rm Cu/Mo}$
and the orbital energy $\varepsilon_{\rm Cu/Mo}$
of the isolated metals depend on their orbital radiuses.
Intermetallic Coulomb, exchange, and pair-hopping interactions
are denoted by $V_{\rm MoCu/CuCu}$,
$J_{\rm MoCu/CuCu}$, and 
$J'_{\rm MoCu/CuCu}$, respectively.
$\displaystyle \sum_{<{\bm n},{\bm m},i,j>}$ runs all Cu-Mo bonds,
while $\displaystyle  \sum_{<{\bm n}\ne{\bm m}>}\sum_{i=1}^{4}$ takes
all pairs of the nearest neighboring Cu sites.

    The model Hamiltonian on the $I4/m$ crystal structure
has the point group symmetry ${\bf C}_{4h}
\equiv$$\{$$E$, $C_{4z}^{-1}$, $C_{2z}$, $C_{4z}$, $\sigma_{h}$,
$IC_{4z}$, $I$, $IC_{4z}^{-1}$$\}$.
We investigate the photoinduced magnetic dynamics
by solving the time-dependent Schr\"odinger equation,
where the electron-electron interactions are treated within
a Hartree-Fock (HF) scheme.
Unless otherwise noted,
the total number of units $N$ is set to $4096=16 \times 16 \times 16$
($\sim$ 50,000 sites).
We confirmed that the static and dynamic properties
of the larger ($N=13,824$) system are
qualitatively the same as those of the $N=4096$ system.
\begin{figure*}[ht]
\centering
\includegraphics[width=140mm]{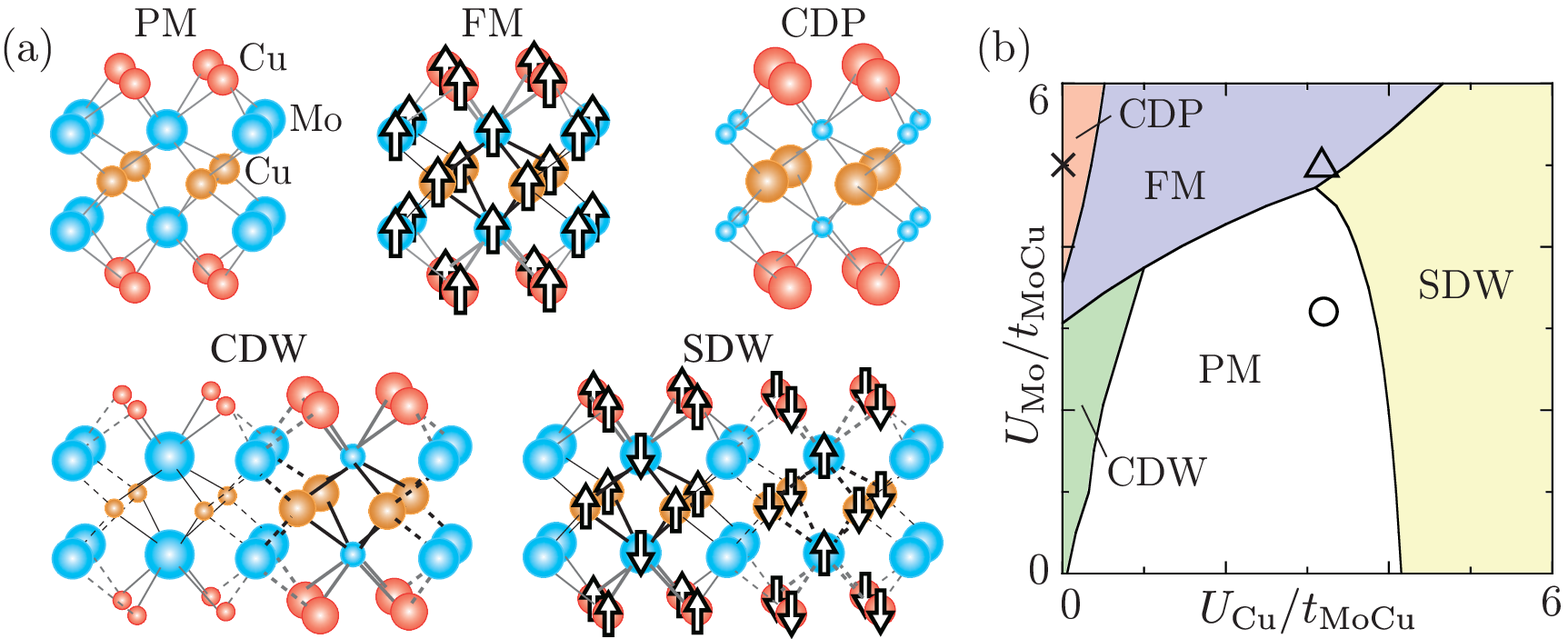}
\caption{
(a) Schematic representations of possible mixed-valent states,
where varying circles and arrows describe
varying electron and spin densities, respectively.
(b) A HF-ground-state phase diagram on
the $U_{\rm Cu}$-$U_{\rm Mo}$ plane,
where the symbols $\circ$ (3.2, 3.2), $\triangle$ (3.2, 5.0),
and $\times$ (0.0, 5.0) indicate the PM, FM, and CDP states,
respectively, in the investigation of photoinduced dynamics.
The other parameters are given as follows,
$t_{\rm CuCu}=0.3t_{\rm MoCu}$,
$V_{\rm CuCu}=2.4t_{\rm MoCu}$,
$V_{\rm MoCu}=0.8t_{\rm MoCu}$,
$\varepsilon_{\rm Cu}-\varepsilon_{\rm Mo}=
1.0t_{\rm MoCu}$, and
$J_{\rm MoCu}=J_{\rm MoCu}'=-J_{\rm CuCu}=-J_{\rm CuCu}'=$
$0.6t_{\rm MoCu}$.}
\label{F:S2}
\end{figure*}

\section*{S2. Competing ground states}
\label{S:S2}
    Figures \ref{F:S2}(a) and \ref{F:S2}(b) show
possible electronic states of the dicopper octacyanomolybdate
and their competition, respectively.
The ground-state phase diagram
of the Hubbard Hamiltonian ${\cal H}_{\rm H}$ [Fig. \ref{F:S2}(b)]
is obtained within the HF scheme,
where the temperature $k_{\rm B}T/t_{\rm MoCu}$
is set to 0.001 in line with the experimental
observation at the sufficiently low temperatures 3 $\sim$ 10 K
\cite{O270}.
The paramagnetic (PM) state having the
full symmetry of the Hamiltonian can take
the valence arrangement
Cu$^{\rm II}$-Mo$^{\rm IV}$-Cu$^{\rm II}$
and predominates in the moderate correlation regime.
This state corresponds to the observed ground state.
There exist two static magnetic solutions,
namely ferromagnetic (FM) and spin-density-wave (SDW) states.
The former shows a spatially uniform electron distribution,
whereas the later exhibits
inter-unit-cell antiferromagnetic ordering
on Mo and Cu sublattices.
In the strong correlation regime,
such magnetic structures are generally stabilized
as is discussed in other correlated electron systems
\cite{H2354,W12749,C104514}.
However, the extremely strong $U_{\rm Mo}$ favors
a nonmagnetic state.
It is the charge-disproportionated (CDP) state
taking Cu$^{\rm I}$-Mo$^{\rm VI}$-Cu$^{\rm I}$
(fully-filled Cu and vacant Mo sites).
Given that the Mo 4$d$ orbital is more
diffuse than the Cu 3$d$ one,
the situation $U_{\rm Mo} >\!\!> U_{\rm Cu}$ is not realistic
and therefore the CDP state hardly appears as a ground state
in the as-grown samples.
In the weak correlation regime,
the charge-density-wave (CDW) state is stabilized
by the intermetallic Coulomb interaction $V_{\rm CuCu}$.

    We focus on the PM state indicated by
$\circ$ in Fig. \ref{F:S3}(b)
and investigate the photoexcitation mechanism.
The model parameters are tuned so as to reproduce
the Cu$^{\rm II}$-Mo$^{\rm IV}$-Cu$^{\rm II}$ structure.
This PM ground state shows a visible-light absorption
[see the inset in Fig. 3(a) in the main text],
which is in good agreement with the observations.
\cite{O270}
Since the PM state has an insulating gap of 0.4 eV
[see Fig. 5(b) in the main text],
there is no thermal fluctuation effect
at the present temperature.
Electronic structures of the FM and CDP states
are also discussed in the main text.
These two states are indicated by symbols
($\triangle$ and $\times$) in the phase diagram.

\section*{S3. Dzyaloshinskii-Moriya and Zeeman interactions}
\label{S:S3}
    A Dzyaloshinskii-Moriya (DM) interaction
is given as
\begin{align}
   &{\cal H}_{\rm DM}=
   \sum_{{\bm n}}\sum_{l=1}^{4}\sum_{\rho=0}^{1}\sum_{\sigma,\sigma'=0}^{1}
   (-1)^{\rho+\sigma}{\bm D}_{l+4\rho}^{(\sigma'\sigma)}
   \nonumber \\ &
   \cdot
   [{\bm s}_{{\bm n}+{\bm \delta(l,\rho,\sigma,\sigma')}:{\rm Mo}(1+3\sigma'+(-1)^{\sigma'}\rho)} \times
    {\bm s}_{{\bm n}:{\rm Cu}(2l-\sigma)}],
   \label{E:DM}
\end{align}
where
${\bm\delta}(l,\rho,\sigma,\sigma')
=\sigma'{\rm Re}[f(l)]({\bm a}/a)
+\sigma'{\rm Im}[f(l)]({\bm b}/a)
+\sigma\rho({\bm c}/c)$ with
$f(l)=e^{{\rm i}\pi/4}[1+{\rm e}^{{\rm i}\pi (1-l)/2}]/\sqrt{2}$.
An electron spin is given as
$s_{{\bm n}:A(i)}^{\lambda}
=\sum_{\sigma,\sigma'}
 c_{{\bm n}:A(i)\sigma}^{\dag}c_{{\bm n}:A(i)\sigma'}
 \tau_{\sigma\sigma'}^\lambda/2$
with the Pauli matrix $\tau^{\lambda}$ $(\lambda=a,b,c)$.
The DM vectors should be compatible with the crystalline structure as
${\bm D}_{i}^{(\sigma'\sigma)}=g_i\cdot{\bm D}_{1}^{(\sigma'\sigma)}$ with
$g_i(\in\mathbf{C}_{4h})
=C_{4z}^{-1}$, $C_{2z}$, $C_{4z}$, $\sigma_{h}$,
$IC_{4z}$, $I$, $IC_{4z}^{-1}$ for $i=2$ to $8$, respectively.
We take 
${\bm D}_{1}^{(\sigma'\sigma)}={\bm D}
=D(\sin\theta,0,\cos\theta)$ with $D=0.7$ meV.
The value of $D$ is comparable to that of typical transition-metal oxides \cite{S775,K3946}.
When the DM vectors are parallel to the ${\bm c}$ (${\bm z}$)
direction ($\theta=0$),
the system is never photomagnetized
because the ${\bm z}$ component of ${\bm S}_{A}\times{\bm S}_{B}$
does not break the conservation law of the total magnetization:
$[{\bm S}_{\rm A}\times{\bm S}_{\rm B}]_{z}
=(S_{\rm A}^{-}S_{\rm B}^{+}-S_{\rm A}^{+}S_{\rm B}^{-})/4i$.
Once ${\bm D}$ has the ${\bm a}$-direction component,
however,
the system can be photomagnetized
even though its in-plane component is very small.
In the main text,
the DM vectors lie on the ${\bm a}$-${\bm b}$ plane
($\theta=\pi/2$).

    We apply an external magnetic field
along the ${\bm c}$ (${\bm z}$) axis to the system
by employing the Zeeman Hamiltonian
\begin{align}
   {\cal H}_{\rm ex}=-g\mu_{\rm B}H\sum_{{\bm n}}
   \Biggl[
      \sum_{i=1}^{8}s^{c}_{{\bm n}:{\rm Cu}(i)}
     +\sum_{j=1}^{4}s^{c}_{{\bm n}:{\rm Mo}(j)}
   \Biggr],
   \label{E:MAG}
\end{align}
where $g\mu_{\rm B}H/t_{\rm MoCu}=2\times10^{-4}$.
The degeneracy of up- and down-spin electrons is lifted,
so that the down-spin electrons may selectively be photoexcited.

We confirmed that the DM and Zeeman interactions
hardly change the phase diagram of ${\cal H}_{\rm H}$ [Fig. \ref{F:S2}(b)]
and have little effect on the static electronic structure of each state
such as a dispersion relation.

\section*{S4. Substitution of tungsten for molybdenum}
\label{S:S4}
    Let us consider substitution of W for Mo
and its effects on the intermetallic transfer integral.
When we apply a half-filled single-orbital Hubbard
model to a two-site system consisting of A and B metals,
the antiferromagnetic interaction 
\begin{align}
J_{\rm AB}=t_{\rm AB}^{2}
\left(\frac{1}{U_{\rm A}}+\frac{1}{U_{\rm B}}\right)
\label{E:JAF}
\end{align}
is obtained in the strong-correlation limit,
where $t_{\rm AB}$ is the A-B transfer integral
and $U_{\rm A/B}$ is the on-site Coulomb repulsion.
We here focus on the octacyanometalate-based
Mn-Mo \cite{W6802} and Mn-W \cite{Y26} antiferromagnets:
$J_{\rm MoMn}=6.9$ cm$^{-1}$
and $J_{\rm WMn}=4.6$ cm$^{-1}$.
Substituting these parameters for Eq.(\ref{E:JAF}),
we obtain
\begin{align}
\frac{t_{\rm WMn}}{t_{\rm MoMn}}
=\sqrt{0.66
\frac{U_{\rm W}+U_{\rm Mn}(U_{\rm W}/U_{\rm Mo})}{U_{\rm W}+U_{\rm Mn}}}.
\end{align}
In the situation $U_{{\rm Mn}(3d)} \ge U_{{\rm Mo}(4d)} \ge U_{{\rm W}(5d)}$,
$t_{\rm WMn}/t_{\rm MoMn}$ is less than unity.
Assuming that $U_{\rm Mo}/U_{\rm Mn}=0.55$ and $U_{\rm W}/U_{\rm Mo}=0.60$,
we obtain $t_{\rm WMn} = 0.65t_{\rm MoMn}$.

    Based on the above consideration,
we set $t_{\rm WCu}$ to 0.42 eV ($=0.65t_{\rm MoCu}$)
in the investigation of the tungsten analog.
The other parameters are
$t_{\rm CuCu}=0.3t_{\rm WCu}$,
$U_{\rm W}=2.0t_{\rm WCu}$,
$U_{\rm Cu}=3.2t_{\rm WCu}$,
$V_{\rm CuCu}=2.4t_{\rm WCu}$,
$V_{\rm WCu}=0.8t_{\rm WCu}$,
$\varepsilon_{\rm Cu}-\varepsilon_{\rm W}=-1.0t_{\rm WCu}$, and
$J_{\rm WCu}=J_{\rm WCu}'=-J_{\rm CuCu}=-J_{\rm CuCu}'=$
$0.6t_{\rm WCu}$.
These parameters are tuned so as to reproduce
the Cu$^{\rm II}$-W$^{\rm IV}$-Cu$^{\rm II}$
structure in the PM ground state.
The obtained absorption spectrum
is also consistent with the observations \cite{R27}
[see the inset in Fig. 3(b) in the main text].

\section*{S5. Polarized optical conductivity}
\label{S:S5}
    The regular part of the optical conductivity, $\sigma_{\lambda}^{\rm reg}(\omega)$
$({\lambda}=a, b, c)$, is defined as
\begin{align}
&   \sigma_{\lambda}^{\rm reg}(\omega)
    =\frac{\pi}{N\omega}\sum_{i\neq 0}
   |\langle E_i|{\cal J}_{\lambda}|E_0\rangle|^2
   \delta(E_i-E_0-\hbar\omega),
   \label{E:OCReg}
\end{align}
where a total unit number is denoted by $N$ 
and the current operator ${\cal J}_{\lambda}$ is
given as
\begin{align}
   &
   {\cal J}_{\lambda}
   ={\rm i}
   \sum_{<{\bm n},{\bm m},i,j>}\sum_{\sigma}
   \frac{e r_{{\rm Cu}(i);M(j)}^{\lambda}}{\hbar}
    \Bigl[(-1)^{i+j+1}t_{M{\rm Cu}}
   \nonumber \\ &
    -V_{M{\rm Cu}} p_{{\bm n}:{\rm Cu}(i);{\bm m}:M(j)}^{\sigma *}
    +J_{M{\rm Cu}} p_{{\bm n}:{\rm Cu}(i);{\bm m}:M(j)}^{*}
   \nonumber \\ &
    +J_{M{\rm Cu}}'p_{{\bm n}:{\rm Cu}(i);{\bm m}:M(j)}^{-\sigma} \Bigr]
    c_{{\bm n}:{\rm Cu}(i)\sigma}^\dagger c_{{\bm m}:M(j)\sigma}
   \nonumber \\ &
   +{\rm i}
   \sum_{<{\bm n}\ne{\bm m}>}\sum_{i=1}^{4}\sum_{\sigma}
   \frac{e r_{{\rm Cu}(i);{\rm Cu}(\bar{i})}^{\lambda}}{\hbar}
    \Bigl[t_{\rm CuCu}
    -V_{\rm CuCu} p_{{\bm n}:{\rm Cu}(i);{\bm m}:{\rm Cu}(\bar{i})}^{\sigma *}
   \nonumber \\ &
    +J_{\rm CuCu} p_{{\bm n}:{\rm Cu}(i);{\bm m}:{\rm Cu}(\bar{i})}^{*}
    +J_{\rm CuCu}'p_{{\bm n}:{\rm Cu}(i);{\bm m}:{\rm Cu}(\bar{i})}^{-\sigma} \Bigr]
   \nonumber \\ & \times
    c_{{\bm n}:{\rm Cu}(i)\sigma}^\dagger c_{{\bm m}:{\rm Cu}(\bar{i})\sigma}
   +{\rm H.c.}
   \hspace{5mm}
   (M={\rm Mo}, {\rm W}).
   \label{E:J}
\end{align}
A bond order is defined as
$p_{{\bm n}:A(i);{\bm n'}:A'(i')}^{\sigma}=
\langle c_{{\bm n}:A(i)\sigma}^{\dagger}c_{{\bm n'}:A'(i')\sigma}\rangle$,
where $\langle \cdots \rangle$ means
a canonical ensemble average,
and a relative vector between $A(i)$ and $A'(i')$ sites
is denoted by ${\bm r}_{A(i);A'(i')}=
\left[r_{A(i);A'(i')}^{a}, r_{A(i);A'(i')}^{b}, r_{A(i);A'(i')}^{c}\right]$.
$|E_i\rangle$ is an arbitrary wavefunction of energy $E_i$
($E_0\leq E_1\leq E_2\leq\cdots$).
The ground state $|E_0\rangle$ and the excited state $|E_i\rangle$ ($i \ne 0$)
are given as
\begin{align}
   |E_0\rangle&
  =\prod_{\varepsilon_{{\bm k},\chi}\leq\varepsilon_{\rm F}}
   \tilde{c}_{{\bm k},\chi}^\dagger|0\rangle
\hspace{3mm} (\text{$|0\rangle$: true electron vacuum}),
\label{E:HFGS}
\\
   |E_i\rangle&=\tilde{c}_{{\bm k},\nu}^\dagger \tilde{c}_{{\bm k},\mu}|E_0\rangle
\hspace{3mm} (\varepsilon_{{\bm k},\mu} \le \varepsilon_{\rm F}< \varepsilon_{{\bm k},\nu}),
\label{E:HFEX}
\end{align}
respectively,
where $\varepsilon_{\rm F}$ is the Fermi energy
and $\tilde{c}_{{\bm k},\chi}^\dagger$
creates an electron of the HF energy $\varepsilon_{{\bm k},\chi}$
($\chi=1\sim24$).

    The Drude component of the optical conductivity, $\sigma_{\lambda}^{\rm D}$,
 is defined as
\begin{align}
&   \sigma_{\lambda}^{\rm D}
  =-\frac{\pi}{N}
    \biggl(
     \langle E_{0}|{\cal T}_{\lambda}|E_{0} \rangle
    +2\sum_{i\neq 0}
      \frac{|\langle E_i|{\cal J}_{\lambda}|E_0\rangle|^2}{E_i-E_0}
    \biggr),
    \label{E:OCreg}
\end{align}
where the kinetic-energy operator ${\cal T}_{\lambda}$ is
given as
\begin{align}
   &
   {\cal T}_{\lambda}
   =\sum_{<{\bm n},{\bm m},i,j>}\sum_{\sigma}
   \Bigl[\frac{e r_{{\rm Cu}(i);M(j)}^{\lambda}}{\hbar}\Bigr]^{2}
   \Bigl[(-1)^{i+j+1}t_{M{\rm Cu}}
   \nonumber \\ &
    -V_{M{\rm Cu}} p_{{\bm n}:{\rm Cu}(i);{\bm m}:M(j)}^{\sigma *}
    +J_{M{\rm Cu}} p_{{\bm n}:{\rm Cu}(i);{\bm m}:M(j)}^{*}
   \nonumber \\ &
    +J_{M{\rm Cu}}'p_{{\bm n}:{\rm Cu}(i);{\bm m}:M(j)}^{-\sigma} \Bigr]
    c_{{\bm n}:{\rm Cu}(i)\sigma}^\dagger c_{{\bm m}:M(j)\sigma}
   \nonumber \\ &
   +\sum_{<{\bm n}\ne{\bm m}>}\sum_{i=1}^{4}\sum_{\sigma}
   \Bigl[\frac{e r_{{\rm Cu}(i);{\rm Cu}(\bar{i})}^{\lambda}}{\hbar}\Bigr]^{2}
    \bigl[t_{\rm CuCu}
    -V_{\rm CuCu} p_{{\bm n}:{\rm Cu}(i);{\bm m}:{\rm Cu}(\bar{i})}^{\sigma *}
   \nonumber \\ &
    +J_{\rm CuCu} p_{{\bm n}:{\rm Cu}(i);{\bm m}:{\rm Cu}(\bar{i})}^{*}
    +J_{\rm CuCu}'p_{{\bm n}:{\rm Cu}(i);{\bm m}:{\rm Cu}(\bar{i})}^{-\sigma} \Bigr]
   \nonumber \\ & \times
    c_{{\bm n}:{\rm Cu}(i)\sigma}^\dagger c_{{\bm m}:{\rm Cu}(\bar{i})\sigma}
   +{\rm H.c.}
   \hspace{5mm}
   (M={\rm Mo}, {\rm W}).
   \label{E:T}
\end{align}
The Drude weight is closely related to the kinetic energy of electrons.
The total optical conductivity is given as
\begin{align}
\sigma_{\lambda}(\omega)=
\sigma_{\lambda}^{\rm reg}(\omega)
+\sigma_{\lambda}^{\rm D}\delta(\omega).
\end{align}

    In the time-evolutional calculation,
the initial state $|E_0 \rangle$ is replaced by
the time-evolutional wavefunction $|\Psi(t) \rangle$.
The one-electron-hole pair excited state from $|\Psi(t) \rangle$ is defined as
\begin{align}
|E_i (t)\rangle=\tilde{c}_{{\bm k},\nu}^\dagger \tilde{c}_{{\bm k},\mu}|\Psi(t)\rangle,
\end{align}
where $\tilde{c}_{{\bm k},\nu}^{\dagger}$ ($\tilde{c}_{{\bm k},\mu}$)
creates (destroys) an electron of the HF-level energy of ${\cal H}_{\rm HF}(t)$.
The regular part and Drude component of
the optical conductivity are given as
\begin{align}
&   \sigma_{\lambda}^{\rm reg}(t; \omega)
    =\frac{\pi}{N\omega}\sum_{i\neq 0}
   |\langle E_i (t)|{\cal J}_{\lambda}|\Psi(t) \rangle|^2
   \delta(\langle E_i \rangle-\langle E_0 \rangle-\hbar\omega)
   \label{E:OCRegt}
\end{align}
and
\begin{align}
&   \sigma_{\lambda}^{\rm D}(t)
  =-\frac{\pi}{N}
    \biggl(
     \langle \Psi(t)|{\cal T}_{\lambda}|\Psi(t) \rangle
    +2\sum_{i\neq 0}
      \frac{|\langle E_i (t)|{\cal J}_{\lambda}|\Psi(t)\rangle|^2}
      {\langle E_i \rangle-\langle E_0 \rangle}
    \biggr),
    \label{E:OCregt}
\end{align}
respectively,
where
\begin{align}
\langle E_0 \rangle=&
\langle \Psi(t) | {\cal H}_{\rm HF}(t) |\Psi(t) \rangle,\\
\langle E_i \rangle=&
\langle E_{i}(t) | {\cal H}_{\rm HF}(t) |E_{i}(t) \rangle.
\end{align}
This is the naivest extension
of Eqs. (\ref{E:OCReg}) and (\ref{E:OCreg}).
A formulation of dynamical correlation functions
in nonequilibrium states is challenging topic
and there are several researches.
One of them treats both pump and probe photons explicitly
and considers inter-excited-state transitions \cite{O085107}.
We confirmed that the visible absorption spectra obtained
with this method are qualitatively the same as those by
Eqs. (\ref{E:OCRegt}) and (\ref{E:OCregt}).

\section*{S6. Parameter dependences of the photoinduced magnetization dynamics}
\begin{figure}[ht]
\centering
\includegraphics[width=60mm]{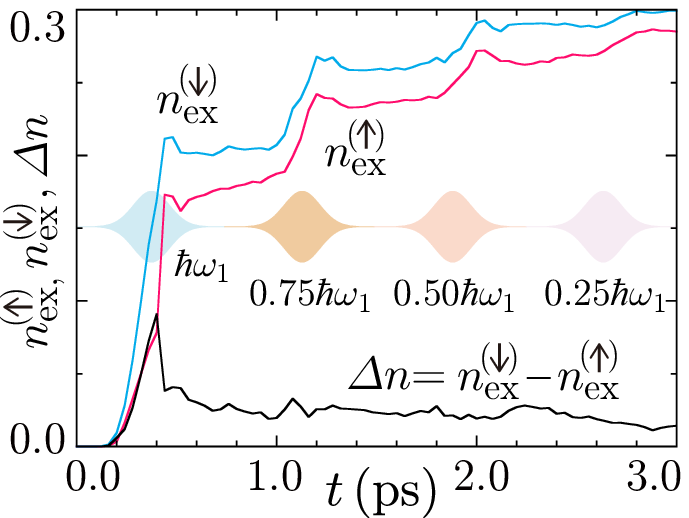}
\caption{
The numbers of excited up-spin ($\uparrow$) and
down-spin ($\downarrow$) electrons per site
in the magnetic phase transition.
The model parameters are the same as
those in Fig. 4(a) in the main text.}
\label{F:S3}
\end{figure}

    In the time-evolutional calculation,
we define the numbers of excited up-spin ($\uparrow$) and
down-spin ($\downarrow$) electrons per site,
\begin{align}
n_{\rm ex}^{(\uparrow)}
&=\frac{1}{12N}\sum_{{\bm k}}
\sum_{\chi=17}^{20}
\langle{\Psi(t)}|
\tilde{c}_{{\bm k},\chi}^{\dagger}\tilde{c}_{{\bm k},\chi}
|{\Psi(t)}\rangle, \\
n_{\rm ex}^{(\downarrow)}
&=\frac{1}{12N}\sum_{{\bm k}}
\sum_{\chi=21}^{24}
\langle{\Psi(t)}|
\tilde{c}_{{\bm k},\chi}^{\dagger}\tilde{c}_{{\bm k},\chi}
|{\Psi(t)}\rangle.
\end{align}
Figure \ref{F:S3} shows
$n_{\rm ex}^{(\uparrow)}$, $n_{\rm ex}^{(\downarrow)}$,
and $\Delta n$ $(=n_{\rm ex}^{(\downarrow)}-n_{\rm ex}^{(\uparrow)})$
as functions of time.
Since the spin degeneracy is lifted due to the external magnetic field,
the higher-lying down-spin electrons are selectively
excited by the first photoirradiation.
This spin-selective electron transfer
induces a net spin density on each site
and then the DM interaction begins to work.
When the system is further irradiated with the longer-wavelength lights,
$\Delta n$ gets smaller.

\begin{figure}[ht]
\centering
\includegraphics[width=86mm]{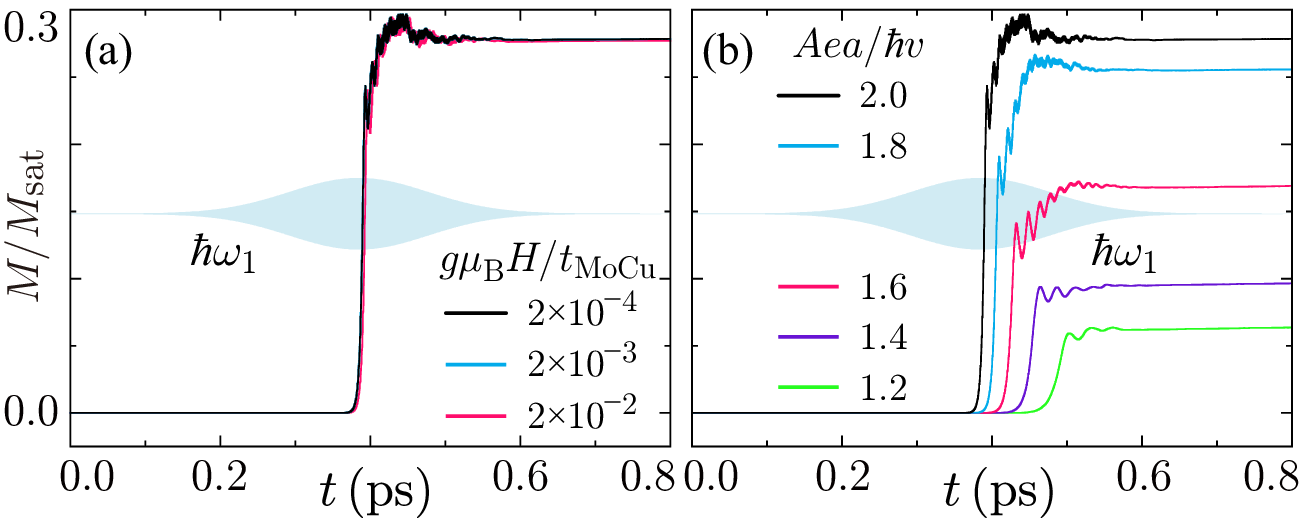}
\vspace*{-5mm}
\caption{
(a) $H$ and (b) $A$ dependences of the magnetization dynamics.
The other parameters are the same as those in Fig. 4(a) in the main text.
The black lines correspond to the result presented in the main text.}
\label{F:S4}
\end{figure}

\begin{figure}[ht]
\centering
\includegraphics[width=86mm]{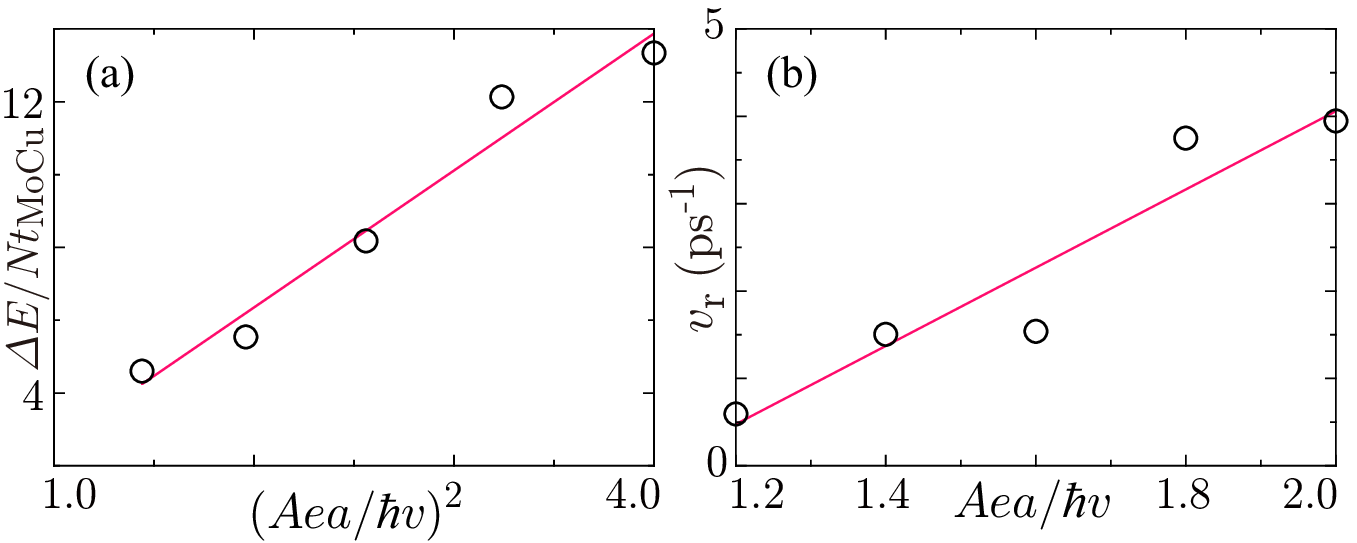}
\vspace*{-5mm}
\caption{
(a) The absorption energy $\Delta E$ at $t=0.8$ ps as a function of the laser power $A^{2}$.
(b) The switching speed $v_{\rm s}$ as a function of the laser amplitude $A$.
Lines are guides to eyes.}
\label{F:S5}
\end{figure}

    Figure \ref{F:S4}(a) shows the magnetic-field ($H$)
dependence of the induced magnetization.
A change in $H$ has little effect on the magnetization dynamics.
In contrast, the magnetization dynamics strongly
depends on the laser amplitude $A$
as is shown in Fig. \ref{F:S4}(b).
When the magnetization rises at time $t_{0}$
and reaches its maximum value $M_{\rm max}$
at time $t_{1}$, the switching speed is defined as
\begin{align}
v_{\rm s}=\frac{M_{\rm max}/M_{\rm sat}}{t_{1}-t_{0}},
\end{align}
where $M_{\rm sat}$ means the saturation magnetization.
The absorption energy $\Delta E$ is given by
\begin{align}
\Delta E=
 \langle \Psi(t) | {\cal H}_{\rm HF}(t) |\Psi(t) \rangle
-\langle \Psi(-\infty) | {\cal H}_{\rm HF}(t) |\Psi(-\infty) \rangle.
\end{align}
In the moderate excitation regime,
$\Delta E$ is proportional to the laser power $A^{2}$,
whereas $v_{\rm s}$ is proportional to the laser amplitude $A$,
as is shown in Fig. \ref{F:S5}.
The amplitude $A$ is decisive for the time scale of the magnetization dynamics.

\begin{figure}[ht]
\centering
\includegraphics[width=86mm]{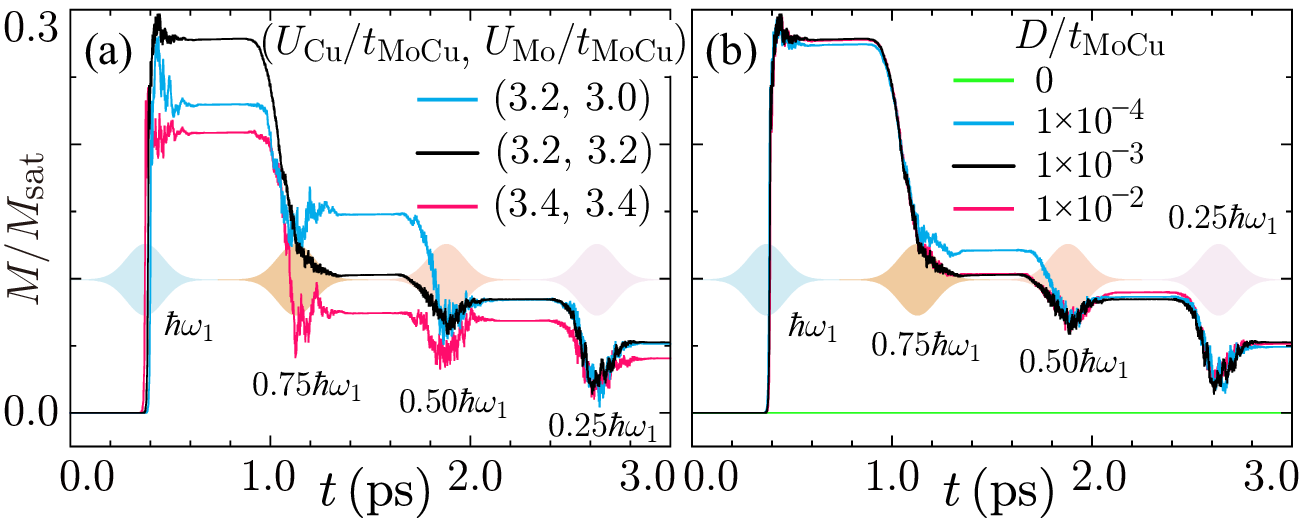}
\vspace*{-5mm}
\caption{
The same as Fig. 4(a) in the main text,
but for different (a) $U_{{\rm Cu}/{\rm Mo}}$ and (b) $D$.
The black lines correspond to the result presented in the main text.}
\label{F:S6}
\end{figure}

    Figure \ref{F:S6}(a) shows
the initial-state dependences of the photoinduced phase transition,
where the on-site Coulomb repulsions are changed within the PM region.
The induced magnetization is quantitatively different
depending on the electronic correlations
but the photoswitching of magnetization is observed in all cases.
Thus, the photoinduced multistep phase transition
occurs at around the present parameter point.
We further present the effects of
the DM interaction in Fig. \ref{F:S6}(b).
When $D=0$,
no magnetization appears because of
the conservation law of the magnetization.
Once $D$ has a finite value,
however,
the system can be photomagnetized even though its value is very small.
The magnetization and demagnetization
dynamics hardly depend on the magnitude of $D$.